\documentclass[english,aps,prper,reprint,showpacs,titlepage,longbibliography,floatfix]{revtex4-1}   

\usepackage[T1]{fontenc}	% should generally be included for better accented-word behavior
\usepackage[latin9]{inputenc}	% should generally be included for better accent behavior
\usepackage{geometry}		% for controlling page margins
\usepackage{graphicx}
\usepackage[above,below]{placeins}	% allows use of \FloatBarrier command to force section barriers
\usepackage{times}
\usepackage{xcolor}

\usepackage{hyperref}  
\hypersetup{colorlinks=true,urlcolor=blue,citecolor=blue,linkcolor=blue}   
\usepackage{enumitem}          % package for handling list formatting
\setlist{nosep}                 % Tightest spacing for lists. `noitemsep` is more relaxed

\begin{document}

\begin{titlepage}

  \title{Toward a valid instrument for measuring physics quantitative literacy}

  \author{Trevor I. Smith}
  \affiliation{Department of Physics \& Astronomy and Department of STEAM Education, Rowan University, 201 Mullica Hill Rd., Glassboro, NJ 08028, USA} 
  \author{Philip Eaton}
  \affiliation{School of Natural Sciences and Mathematics, Stockton University, Galloway, NJ 08205, USA}
  \author{Suzanne White Brahmia}
  \affiliation{Department of Physics, University of Washington, Box 351560, Seattle, WA 98195-1560, USA}
  \author{Alexis Olsho}
  \affiliation{Department of Physics, University of Washington, Box 351560, Seattle, WA 98195-1560, USA}
  \author{Andrew Boudreaux}
  \affiliation{Department of Physics \& Astronomy, Western Washington University, 516 High St., Bellingham, WA 98225, USA}
  \author{Charlotte Zimmerman}
  \affiliation{Department of Physics, University of Washington, Box 351560, Seattle, WA 98195-1560, USA} 

  % \keywords{}

  \begin{abstract}
    We have developed the Physics Inventory of Quantitative Literacy (PIQL) as a tool to measure students' quantitative literacy in the context of introductory physics topics. We present the results from various quantitative analyses used to establish the validity of both the individual items and the PIQL as a whole. We show how examining the results from classical test theory analyses, factor analysis, and item response curves informed decisions regarding the inclusion, removal, or modification of items. We also discuss how the choice to include multiple-choice/multiple-response items has informed both our choices for analyses and the interpretations of their results. We are confident that the most recent version of the PIQL is a valid and reliable instrument for measuring students' physics quantitative literacy in calculus-based introductory physics courses at our primary research site. More data are needed to establish its validity for use at other institutions and in other courses.
    \clearpage
  \end{abstract}
  %% Adding the `\clearpage` is the hack to make the title page.  In 2020, the proceedings is
  %% going to be double blind.  This change makes it so that we can programmatically remove the
  %% title page.  In the future, other blinding measures should be taken as well (for example,
  %% removing self-citations).  This is not needed in 2019.

  \maketitle
\end{titlepage}

\section{Introduction}
Physics Quantitative Literacy (PQL) is defined as the interconnected skills, attitudes, and habits of mind that together support the sophisticated use of elementary mathematics in the context of physics \cite{thompson2010,ojose2011,White2020neg,Olsho2019b}. Developing PQL is a desired outcome of physics instruction, but valid measures of reasoning about quantities and their relationships in physics contexts are absent from research-based assessment instruments in introductory physics. We have developed the Physics Inventory of Quantitative Literacy (PIQL) to address this need \cite{Olsho2019b}. The PIQL is a reasoning inventory that probes the quantification typically used in introductory physics that has a potential impact analogous to the early concept inventories in physics education research that catalyzed curriculum development efforts by raising awareness of broad instructional goals that are not being met \cite{Hake1998,VonKorff2016,Madsen2017}.

In introductory physics, PQL involves using simple mathematics in sophisticated ways. Reasoning about ratios and proportions, covariation, and signed quantities/negativity are at the heart of quantification in introductory physics \cite{thompson2014,thompson2010,thompson2003}. The PIQL was designed based on these three facets of quantification, with many items being drawn from previous research in mathematics and physics education \cite{carlson2010precalculus,boudreaux2015,brahmia2015,brahmia2016a,Brahmia2017c,brahmia2016b,brahmia2017a}. %Previous results have shown that these three facets may not be identifiable in students' responses to PIQL items, suggesting that the PIQL is testing the single construct of PQL.

\begin{figure*}[bt]
    \includegraphics[width = 0.8\columnwidth]{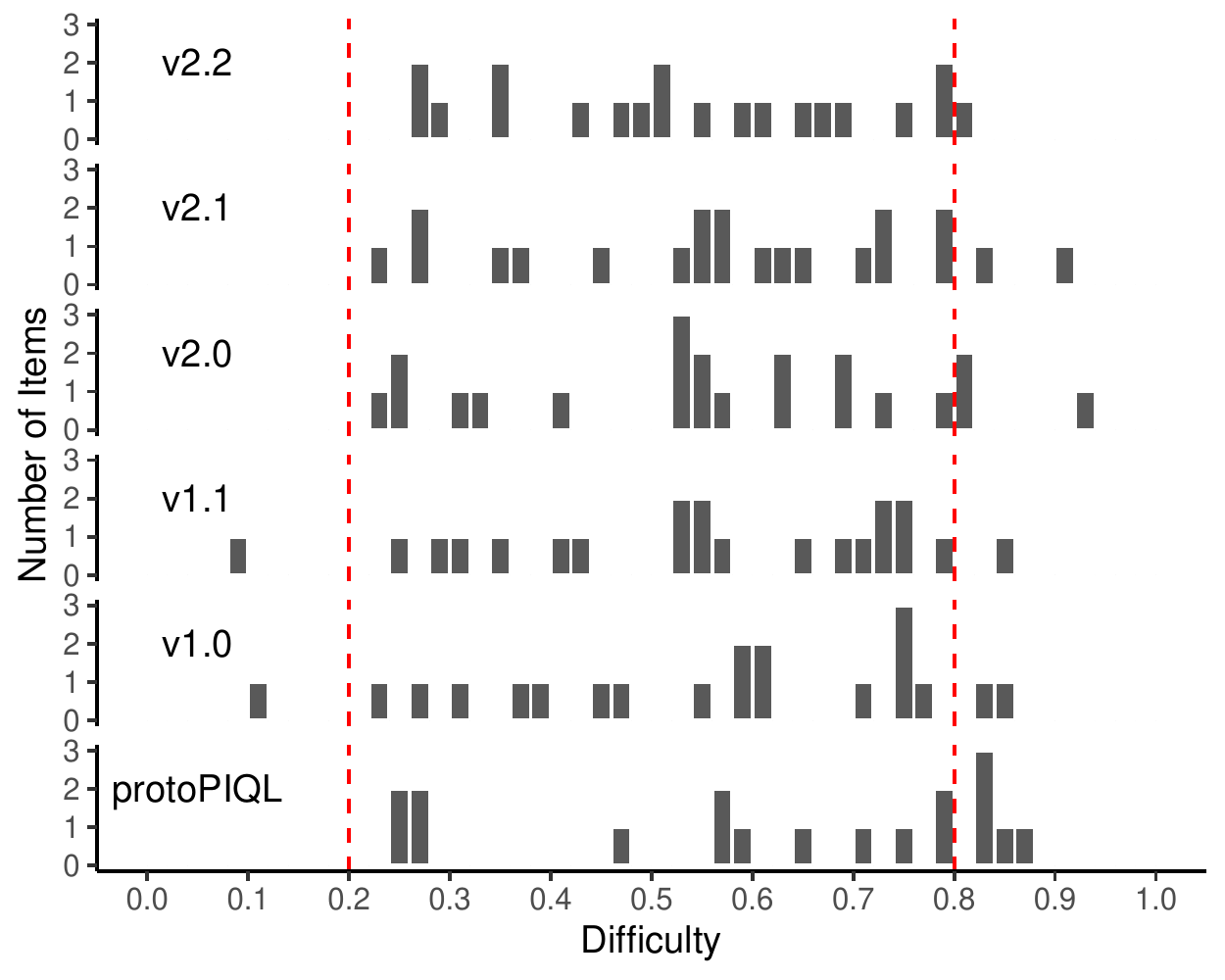}
    ~~~~~~~~
    \includegraphics[width = 0.8\columnwidth]{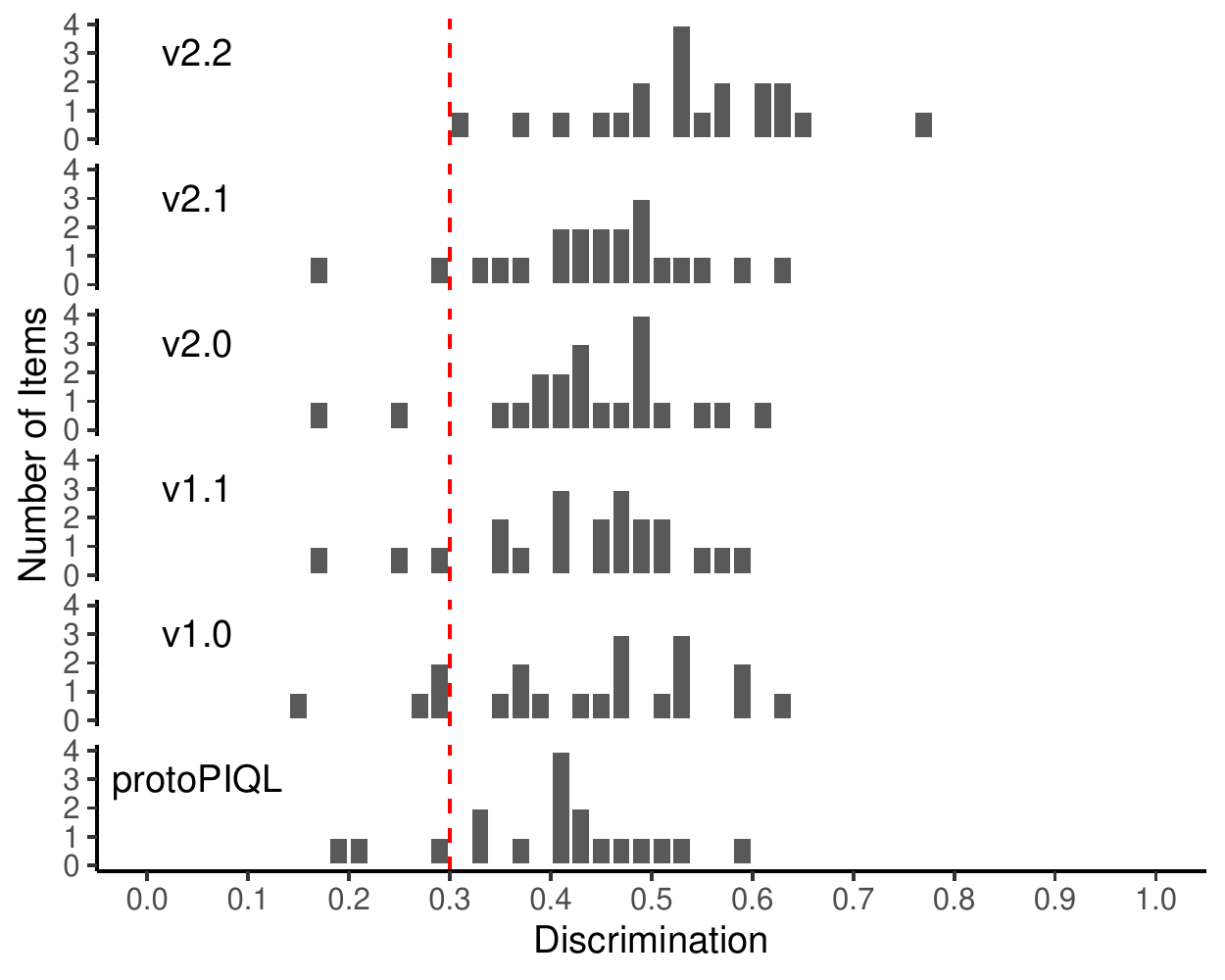}
    
    ~~~~~~(a) \hspace{0.83\columnwidth} (b)
    \caption{CTT difficulty (a) and discrimination (b) parameter distributions for all versions of the PIQL. The desired range of difficulty values is between 0.2 and 0.8 (shown by dashed red lines). The desired range for discrimination is above 0.3.}% Revisions have resulted in more items within the desired range and a greater variety of difficulty values than the prototype version.}
    \label{fig:CTT}
\end{figure*}

% \FloatBarrier

Over the past two years we have engaged in an iterative process of data collection and analysis, item development, and test revision to establish the validity of the PIQL for use in calculus-based introductory physics courses. In the following sections we discuss the methods we used to analyze the data, how we interpreted the results, and the decisions we made to improve the PIQL. Our focus is on using a variety of quantitative measures to gain a holistic view of the PIQL to optimize our ability to measure student reasoning. We present the results from each analysis individually, but our decisions to modify individual items and the PIQL as a whole were based on the collective results from all of them.

\section{Inventory Development and Data Sources}
The prototype version of the PIQL focused primarily on measuring students reasoning about ratios and proportions \cite{brahmia2015,brahmia2016a,Brahmia2017c} and signed quantities/negativity \cite{brahmia2016b,brahmia2017a,White2018,White2019rume}. This 18-item protoPIQL also included two items on covariation taken (with permission) from the Precalculus Conceptual Assessment (PCA) \cite{carlson2010precalculus}. Revisions were made to improve the validity and reliability of the PIQL, reduce redundancies, and ensure that all three facets of PQL were represented. Later versions of the PIQL include 20 or 21 items.  

Data for this study were collected at the beginning of each term (before instruction) in three calculus-based introductory physics courses at a large public research university in the northwestern US. Previous results have shown that overall score distributions on the PIQL are not significantly different in the three courses \cite{Smith2018,Smith2019rume}, and this trend has persisted throughout all versions of the PIQL; therefore, we have combined all data collected in each term for this study. Due to our iterative revisions, the items on the PIQL in each of the six data sets are slightly different; we label the data sets by their version of the PIQL: protoPIQL, v1.0, v1.1, v2.0, v2.1, and v2.2. Data were collected from approximately 1000 students for each version.

\section{Quantitative validation using Classical Test Theory}
We used various quantitative analyses to measure the validity and reliability of each version of the PIQL. Using Classical Test Theory (CTT) we calculated the difficulty and discrimination parameters for each item; we want to have a wide range of difficulty values with most items between 0.2 and 0.8 (representing the fraction of students who answer each item correctly), and we want most discrimination values to be above 0.3 (representing the difference in CTT difficulty between the top and bottom 27\% of students) \cite{Wiersma1990}. We also calculated Cronbach's $\alpha$ as a measure of reliability; a value of at least 0.7 indicates that the test is reliable for measuring the performance of groups of students on a single-construct test, and a value of at least 0.8 indicates that the test is reliable for measuring the performance of individual students \cite{Doran1980}.

Figure \ref{fig:CTT} shows the distributions of the CTT difficulty and discrimination parameters for each version of the PIQL. Five of the items in the protoPIQL were considered too easy (difficulty above 0.8), and three items had discrimination values below 0.3; moreover, there was a gap in the middle of the difficulty distribution with only one item having a difficulty in the range between 0.3 and 0.55. Due to these results, we chose to use only nine of these items in subsequent versions of the PIQL, with one of them being periodically modified. For PIQL v1.0, 11 items were added based on previous research on all three of our PQL facets \cite{brahmia2015,brahmia2016a,Brahmia2017c,brahmia2016b,brahmia2017a,White2018,White2019rume,carlson2010precalculus}, which resulted in a much broader distribution of CTT difficulty values. One additional proportional reasoning item was added to PIQL v1.1; for PIQL v2.0 two covariation items were replaced by newly developed items based on research in mathematics education \cite{Moore2013,Hobson2017,Paoletti2017}; two items were slightly modified for v2.1; one item was removed for PIQL v2.2 due to consistently high difficulty and low discrimintation parameters. 

Taken together, these revisions have resulted in a 20-item instrument with a broad range of difficulty values (only one of which is above the desired upper limit of 0.8), and all items having discrimination values above 0.3. Six of the 20 having large discrimination (above 0.6), meaning that high-scoring students are much more likely to answer these questions correctly than low-scoring students. Additionally, Cronbach's $\alpha$ has also increased: $\alpha = 0.67$  on the protoPIQL, which does not meet the threshold for measuring either groups of students or individuals; however, $\alpha = 0.80$ on PIQL v2.2, which meets both thresholds. The distribution of difficulty values for PIQL v2.2 is a little higher than we think would be ideal (average of 0.54), but we have chosen to keep some of the easier items because we recognize that the students in our data set may have had more prior exposure to mathematics and physics instruction than is typical of the introductory physics student population \cite{kanim2017}. We consider the changes in parameter values to indicate that we have created a valid and reliable inventory for measuring PQL for students in calculus-based introductory physics courses.

\section{Analyzing Data from Multiple-Choice/ Multiple-Response Items}
\label{sec:mcmr}
We consider PQL to be a conceptual blend between physics concepts and mathematical reasoning \cite{Fauconnier2002,White2020rume}. In order to measure the complexity of ideas that students bring from both of these input spaces, we have chosen to include some multiple-choice/multiple-response (MCMR) items in which students are instructed to ``select all statements that \textbf{must be true}'' from a given list, and to ``\textbf{\textit{choose all that apply}}'' (emphasis in the original text). The MCMR item format has the potential to reveal more information about students' thinking than standard single-response items, but it also poses problems with data analysis, as typical analyses of multiple-choice tests (such as CTT) assume single-response items.

% \subsection{Four-level scoring scale} 
For MCMR items, dichotomous scoring methods require a student to choose \emph{all} correct responses and \emph{only} correct responses to be considered correct. For example, item 18 on PIQL v2.2 has two correct answer choices: D and G. In a dichotomous scoring scheme a student who picks only answer D would be scored the same way as a student who chooses answers E and F (incorrect). This ignores the nuance and complexity of students' response patterns within (and between) items. As such, the CTT results for these items are not entirely representative of students' responses.

\begin{figure}[b]
    % \vspace{-6mm}
    \includegraphics[width = 0.45\textwidth]{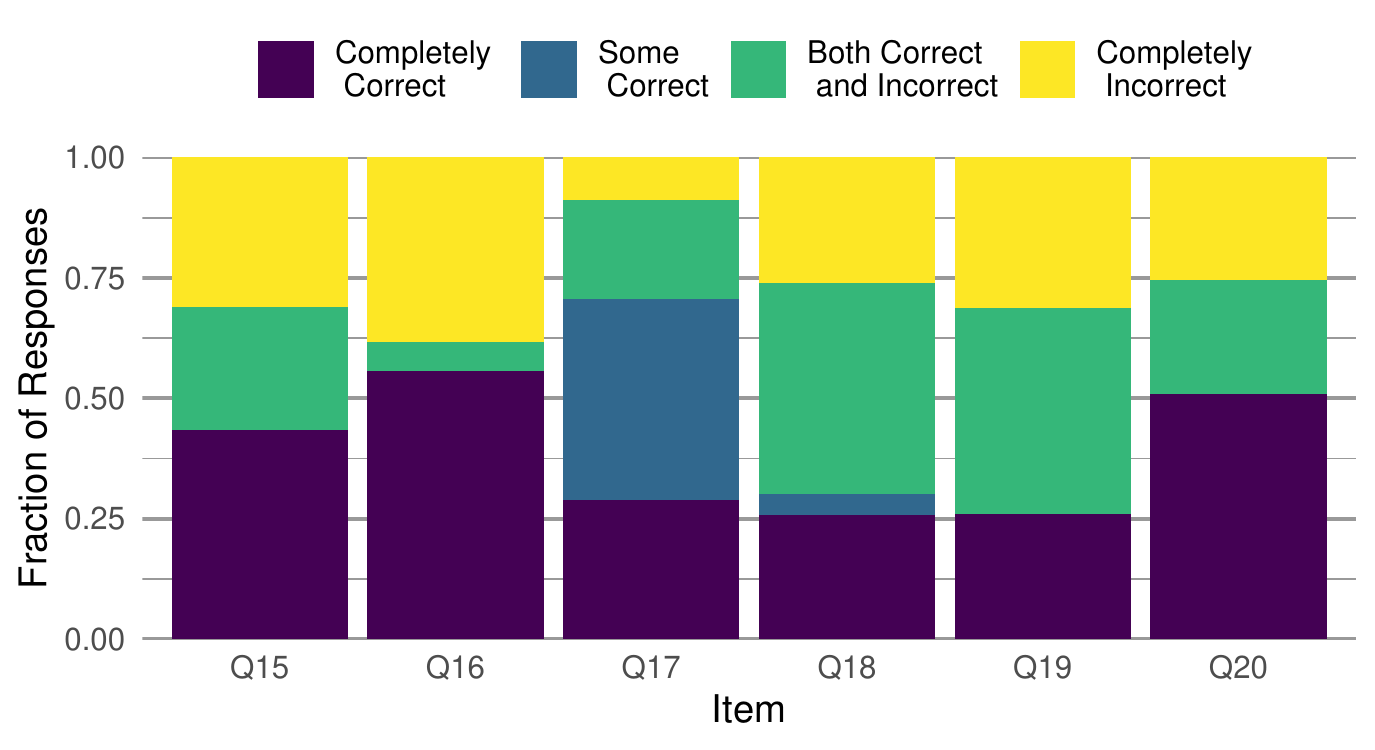}
    \caption{Fraction of student responses in each category of our four-level scoring scheme for MCMR items with multiple correct answers. These results are from the final version of the PIQL.}
    
    \vspace{-3mm}
    \label{fig:correctness}
\end{figure}

In an effort to move beyond the constraints of dichotomous scoring for MCMR items, we have developed a four-level scoring scale in which we categorize students' responses as Completely Correct, Some Correct (if at least one but not all correct response choices are chosen), Both Correct and Incorrect (if at least one correct and one incorrect response choices are chosen), and Completely Incorrect \cite{Smith2018, Smith2019rume}. Figure \ref{fig:correctness} shows the results of using this four-level scoring scale to categorize student responses to the six MCMR items on PIQL v2.2. The dark purple Completely Correct bars are equivalent to CTT difficulty; however, Fig.\ \ref{fig:correctness} also shows us that at least 60\% of students provide at least one correct response to each item (Completely Correct, Some Correct, and Both Correct and Incorrect combined), although this is often coupled with an incorrect response (6\%--44\% of students categorized as Both Correct and Incorrect). This tells a very different story than the CTT results, which group the Some Correct, Both, and Completely Incorrect categories together into a broad Incorrect category.

\begin{figure*}[bt]
    \centering
    \includegraphics[width = 0.3\textwidth]{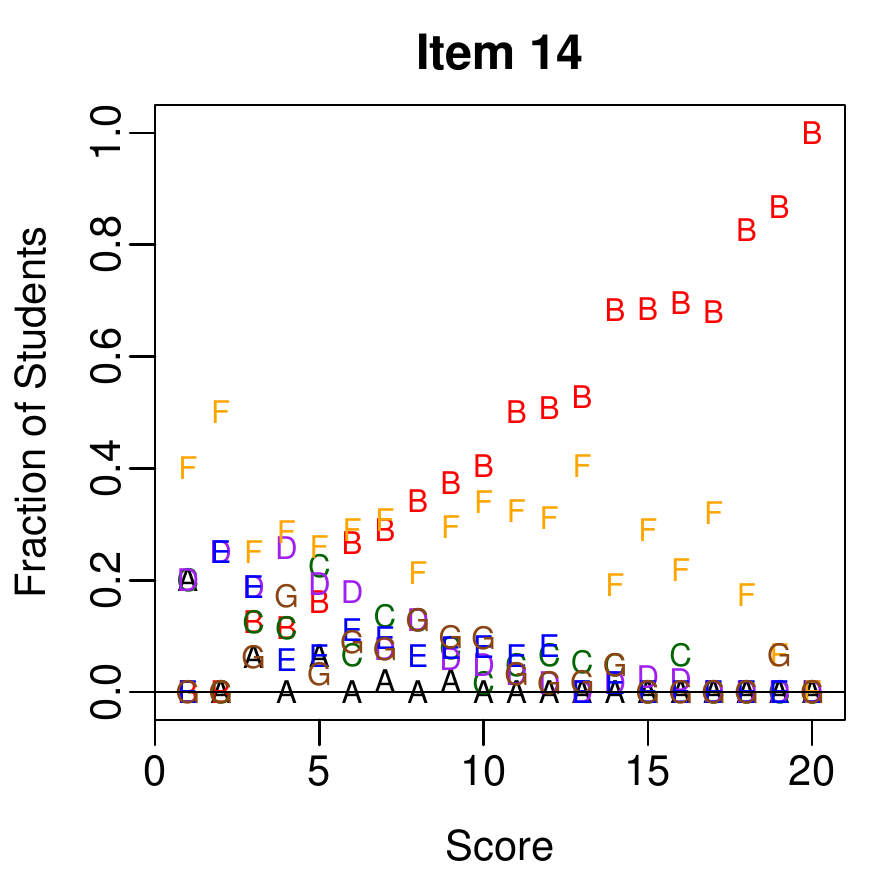} ~~~~ \includegraphics[width = 0.3\textwidth]{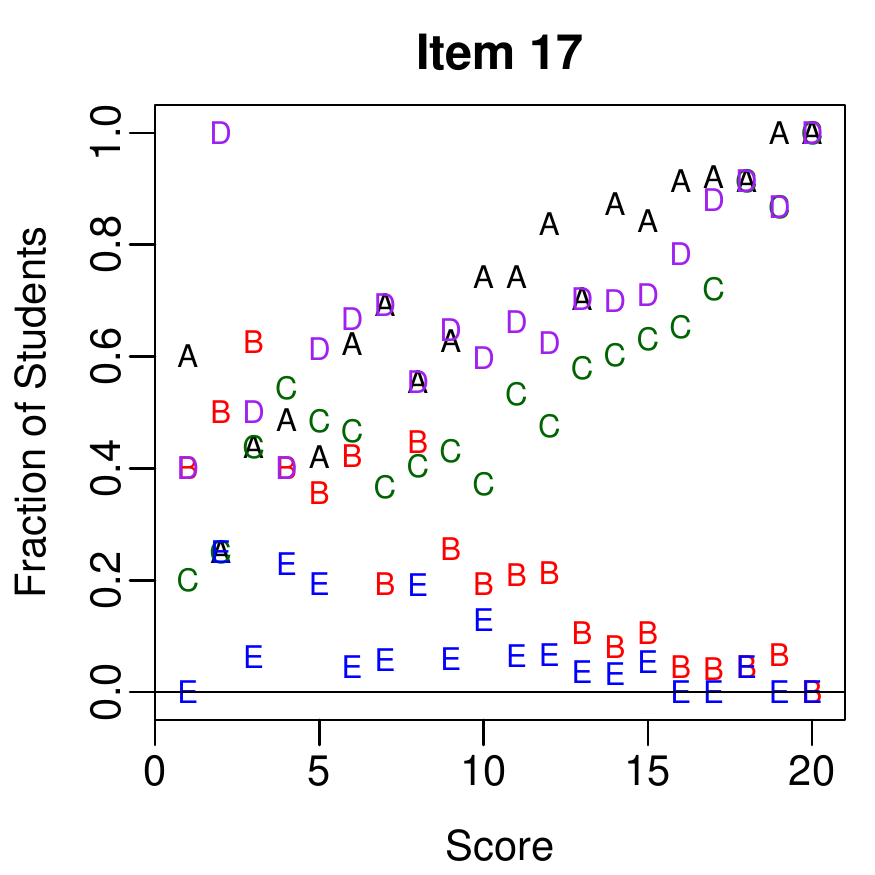} ~~~~ \includegraphics[width = 0.3\textwidth]{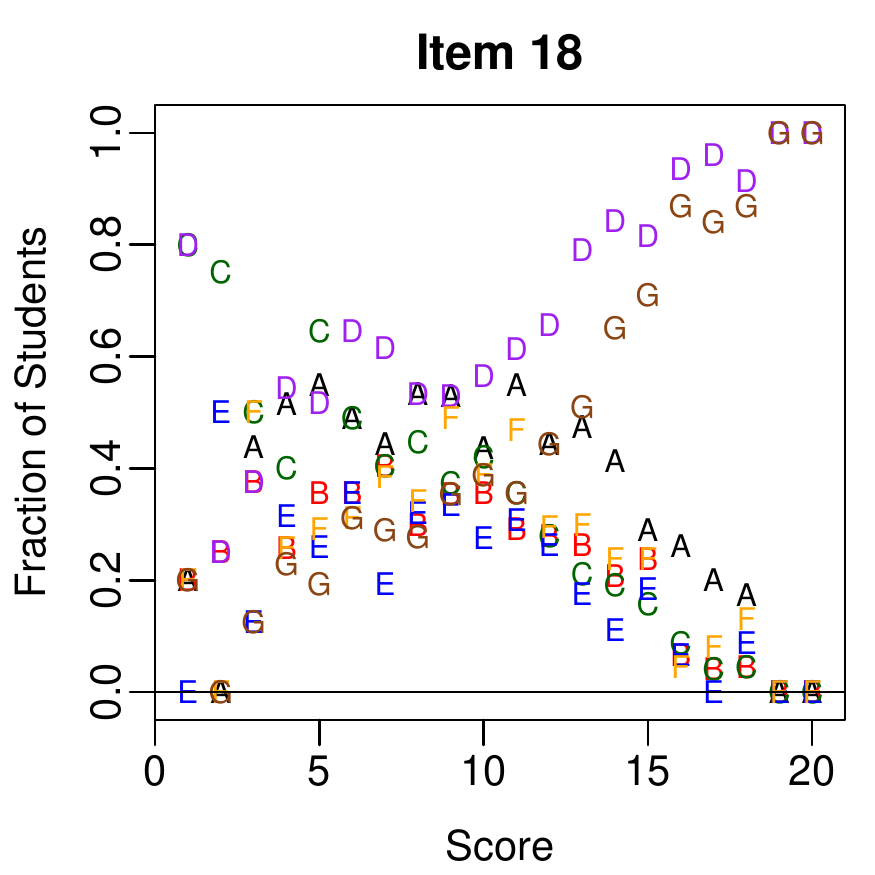}
    \caption{Item Response Curves for three items on PIQL v2.2. Each plot shows the fraction of students who chose each response out of the students who earned each score on the total test. Item 14 has correct answer B, item 17 has correct answers A, C, and D, and item 18 has correct answers D and G.}
    \label{fig:ircs}
\end{figure*}

These four-level scoring results also reveal differences hidden by dichotomous scoring. For example, on PIQL v2.2 two items (Q17 and Q18) have more than one correct answer choice. Figure \ref{fig:correctness} shows that approximately the same number of students answers these items completely correctly, but Q17 has a much higher fraction of students in the Some Correct category. Students are much more likely to include one of the incorrect responses to Q18 than they are for Q17. The items with multiple correct answers also present a new question: is it better for a student to choose Some Correct answers or Both Correct and Incorrect answers? The answer may depend on the specifics of each item and the associated answer choices.%; moreover, different combinations of Both Correct and Incorrect answer choices may be considered better than others. 
Future work will include analyzing data from MCMR items to develop a more sophisticated scoring scheme.

% \subsection{Item Response Curves} 
% \label{sec:irc}

To further examine the responses students give to individual PIQL items we use Item Response Curves (IRCs), which show the fraction of students who choose each answer choice as a function of the students' overall score on the PIQL \cite{Morris2006,Morris2012,Walter2016,Ishimoto2017}. IRCs have been used with single-response tests to rank incorrect responses and to compare different student populations with regard to both correct and incorrect answer choices \cite{Walter2016,Ishimoto2017}. We find IRCs particularly helpful for examining student responses to items with multiple correct answers. 

Figure \ref{fig:ircs} shows three IRCs with different behavior. Item 14 is a single-response item with correct answer B. Even fairly high-scoring students persist in choosing a particular incorrect answer F. Item 17 has three correct responses (A, C, D), with A being the most commonly chosen, and C being the least commonly chosen. Few students at any score level choose E, and fewer than 20\% of students who score above average (10.8) choose either incorrect response (B, E). Item 18 is particularly interesting in that all responses are chosen by 20\%--60\% of students in the middle score range (8-12). This supports the results from Fig.\ \ref{fig:correctness} that students are likely to choose both a correct and an incorrect response to Q18.

Both the four-level scoring scheme and the IRCs provide more information than traditional CTT analyses and allow us to see patterns in students' responses that go beyond typical dichotomous scoring methods. We have used these to gain a deeper qualitative picture of student performance on each PIQL item, and these have been very valuable for deciding which items to keep, eliminate, or modify.

\section{Exploring the substructure of the PIQL}
\label{sec:substructure}

The PIQL was initially developed to probe student reasoning about ratios and proportions, covariation, and signed quantities/negativity. In the language of factor analysis, this would imply that the PIQL was originally intended to have a three-factor structure. Since the intended factor structure of the PIQL was well understood at the beginning of its development, confirmatory factor analysis (CFA) was used at the onset, in conjunction with exploratory factor analysis (EFA). CFA is a model-driven statistical method whose goal is to identify the adequacy of a proposed factor model to response data from the instrument being analyzed \cite{Brown2015}. EFA is a data-driven statistical method whose goal is to uncover the underlying dependencies between observed variables \cite{Lawley1963}. For all versions of the PIQL, CFA determined that the proposed, facet-driven, factor model was not an adequate representation of the PIQL's latent trait structure \cite{Smith2019perc}. The target threshold for CFA is to have goodness-of-fit statistics such as the Confirmatory Fit Index (CFI) and Tucker-Lewis Fit Index (TLI) above a threshold of 0.9 \cite{Eaton2018}. For all versions of the PIQL the CFI and TLI were below 0.8 when using the facet-driven factor model.

Given that the CFA results do not fit with the proposed mode, we moved on to a more in-depth investigation using EFA. The goal of using EFA was to determine if the PIQL has any substructure, and how closely any substructure aligns with the three facets of PQL. The results from parallel analysis suggested that 3--4 meaningful factors could be extracted for the earlier versions of the PIQL (protoPIQL, v1.0, and v1.1) \cite{Weng2005}; however, when examining these structures, they were found to be inconsistent with the originally intended factors, based on the three facets of PQL \cite{Smith2019perc}. During this initial development of the PIQL, EFA models of versions v1.0 and v1.1 each contained a factor that only contained the same two items. % (v1.0: Q11 and Q12; v1.1: Q8 and Q18). 
These two items were found to have item loadings on the same factor of above 0.8, compared to the next highest loading value of approximately 0.5. These items' loadings remained essentially the same when they appeared sequentially on v1.0 and when they were separated and placed onto different pages of the instrument in v1.1. This suggested these items were redundant, %and only one needed to appear on the instrument, 
which lead to the removal of one of the items from the PIQL in future iterations.

Analyses of the most recent versions of the PIQL (v2.0, v2.1, and v2.2) suggest the instrument is now unidimensional, with no strong substructure amongst the items. Results from EFA parallel analysis suggested that these versions of the PIQL could be adequately described by a single factor. Additional evidence to support this conclusion was obtained by performing CFA on v2.1 and v2.2 of the PIQL using a unidimensional model, with measures of goodness-of-fit suggesting that the unidimensional model adequately fit the student response data. Specifically, the CFI and TLI were above 0.93 for both versions under CFA using a unidimensional model. Additionally, the standardized root mean square of the residuals was below 0.04, and the root mean square of the error of approximation was below 0.04 \cite{Eaton2018}. This suggests that removing one of the redundant items identified in v1.0 and v1.1, resulted in the collapse of the PIQL's multiple factor structure into one that is unidimensional. %It is worth noting that from v2.0 to v2.1, two items on the PIQL were replaced. However, these changes did not result in more factors manifesting in v2.1. 
This may also have been affected by replacing two of the covariation items from v1.1.

A major confounding feature of these results is that the factor loadings were determined based on dichotomously scored items. As shown in Fig.\ \ref{fig:correctness}, up to 65\% of students who choose correct responses to MCMR items may be scored as incorrect because either they didn't choose all of the correct responses or they also chose an incorrect response. As such, the factor loadings may not accurately capture the relationships between students' responses for cases involving MCMR items.

To preserve the nuance and complexity of students' response patterns within (and between) items we used module analysis for multiple-choice responses to examine the network of student responses to PIQL items \cite{Brewe2016}. Module analysis uses community detection algorithms to identify modules (a.k.a.\ communities, etc.) within networks of responses to multiple-choice items. We chose to analyze a network of only correct responses to PIQL items. The benefit of this method is that we can examine the patterns that arise from students' selections of each individual correct response, which preserves some of the complexity of MCMR items.

Earlier module analyses of v1.0 and v1.1 using various community detection algorithms on full data sets suggested that there was some substructure in the PIQL. Again, these results did not agree with the three facets that the PIQL was intended to measure and also did not align well with the results of EFA \cite{Smith2019perc, Smith2020rume}. Recent developments in the application of module analysis within PER have enabled a deeper and more refined analysis of the module structure of the PIQL \cite{Wells2019}. Using Modified Module Analysis (MMA) on the final two versions of the PIQL, with a locally adaptive network sparsification (LANS) in place of a global cutoff sparsification, resulted in no discernible substructure between the items on the instrument \cite{Wells2019, Foti2011}. This corroborates the conclusions of EFA and CFA that the PIQL is not measuring multiple constructions and is thus a unidimensional instrument.

\section{Conclusions}
Our goal is to develop a valid and reliable instrument to measure PQL for students in calculus-based introductory physics courses. Results from classical test theory show that after several revisions the items on the PIQL have a broad range of difficulty values, and all items have acceptable levels of discrimination. The reliability of the PIQL has been established with Cronbach's $\alpha = 0.80$, which meets the typically accepted criterion for measuring both properties of groups and properties of individuals.

Results from exploratory and confirmatory factor analysis and modified module analysis show that the PIQL is a unidimensional instrument that measures a single construct. We interpret this construct as being Physics Quantitative Literacy. These results show that student responses to PIQL items do not separate cleanly along the lines of ratios and proportions, covariation, and signed quantities/negativity, suggesting that these three facets of PQL (which are discernible to experts) may develop simultaneously in students.

We have supplemented rigorous psychometric analyses with four-level scoring methods for MCMR items and IRCs, which provide additional information about students' choices of both correct and incorrect responses. These analyses played a vital role in informing our decisions when revising the PIQL. Future work will include developing more sophisticated analyses that can include the nuance of MCMR data into CTT-style analyses.

Additional manuscripts will detail the work we have done to qualitatively validate both individual PIQL items and the inventory as a whole using interviews of both students and experts. As a result of all our analyses, we are comfortable asserting that the PIQL is a valid and reliable instrument for use in calculus-based introductory physics courses at our primary research site. Our next steps will include establishing its validity in broader contexts by collecting data from students in calculus-based introductory courses at other institutions, as well as from students in algebra-based and conceptual physics courses, in order to increase the potential of the PIQL to catalyze meaningful curriculum development efforts.

\acknowledgments{
This work is supported by the National Science Foundation under awards DUE-1832836, DUE-1832880, DUE-1833050, and DGE-1762114.
}

%\clearpage
% For a longer bibliography, delete the thebibliography block above, then comment in these two
% lines to use a .bib file with BibTeX.
% \bibliographystyle{apsrev} % supercedes the longbibliography option, so leave commented out if you want to display article titles
\bibliography{references.bib} % include the .bib suffix in overleaf only

\end{document}